\documentstyle[prl,aps,preprint,tighten]{revtex}
\begin{document}

\draft
\title{Disappearance of the Spin Gap in a Zn doped 2-Leg Ladder Compound
 Sr(Cu$_{1-x}$Zn$_{x}$)$_{2}$O$_{3}$}
\author{M. Azuma and M. Takano}
\address{Institute for Chemical Research, Kyoto University,
Uji, Kyoto-fu 611, Japan}
\author{R. S. Eccleston}
\address{ISIS Facility, Rutherford Appleton Laboratory, Chilton, Didcot, 
Oxfordshire OX10 OQX, United Kingdom}
\date{\today}
\maketitle
\begin{abstract}
An inelastic neutron scattering study was performed on a Zn-substituted 
spin-1/2 Heisenberg 2-leg ladder compound 
Sr(Cu$_{1-x}$Zn$_{x}$)$_{2}$O$_{3}$ ($x \leq 0.04$) to investigate 
nonmagnetic impurity effects on the quantum spin system with a large 
spin gap of about $\sim$ 400 K.  
The magnitude of the spin gap was found to be independent of Zn 
concentration, 33 meV, while the integrated magnetic scattering which corresponds 
to the singlet-triplet excitation decreased monotonically with increasing $x$.
On account of the total-sum rule, this result supports the existence of a finite in-gap density of 
states at $E$ = 0 which was suggested from the $T$-linear magnetic specific heat.
\end{abstract}


Recently, one dimensional antiferromagnets which with gapped singlet spin 
states have attracted much attention. 
Such singlet ground states are realized in three classes 
of Heisenberg 1D antiferromagnetic (AF) systems, 
a $S$ = 1/2 alternate chain found in a 
spin-Peierls system below the transition temperature, a uniform 
integer-spin chain (Haldane material), and a $S$ = 1/2 two leg spin ladder 
compound \cite{Dagotto:Rice}.
SrCu$_{2}$O$_{3}$ \cite{Hiroi} is the most typical compound comprising
 such 2-leg ladder lattice.
In this compound, the ladders made of AF Cu-O-Cu linear 
bonds are connected with each other spatially so that they form 2D 
Cu$_{2}$O$_{3}$ sheets, but the ladders are separated from each 
other magnetically because of the inter-ladder 90$^{{\rm o}}$ Cu-O-Cu bond 
causing spin frustration at the interface.  
The existence of a wide spin gap was found through the measurements
of magnetic susceptibility \cite{Azuma} and nuclear spin relaxation time, 
$T_{1}$  \cite{Ishida}. 
However the former measurement gave the 
magnitude of a gap magnitude of 420 K, whereas that estimated from the 
temperature dependence of $T_{1}$ was 680 K.
This apparent discrepancy was explained theoretically by Kishine and 
Fukuyama based on a Majonara fermion representation \cite{Kishine}, and 
they showed that both experimental data could be explained assuming a gap 
of 440 K.   
Nevertheless, a neutron scattering study has been required to determine the 
magnitude of the spin gap directly.

Nonmagnetic impurities introduced into quantum antiferromagnets have been 
expected to affect the magnetic properties in various ways.  
Of particular interest is the coexistence of lattice dimerization and long-range 
AF ordering in an impurity-substituted spin-Peierls material, 
Cu$_{1-x}$Zn$_{x}$GeO$_{3}$ \cite{Oseroff,Hase2}.  
The coexistence of the two seemingly exclusive phenomena was observed also in 
CuGe$_{0.993}$Si$_{0.007}$O$_{3}$ \cite{Renault,Fukuyama} in which the 
Cu-sublattice was kept clean.  
We have studied the nonmagnetic impurity effects on a 2-leg ladder 
compound through the measurements of magnetic susceptibility and  
specific heat of Sr(Cu$_{1-x}$Zn$_{x}$)$_{2}$O$_{3}$ \cite{Azuma2}.  
The influence of 
the Zn substitution was found to be much more extended than naively 
expected : Instead of the creation 
of free localized spin-1/2's studding the matrix in its singlet state
 \cite{White}, an antiferromagnetic transition 
 at an $x$-dependent temperature below 10 K was seen in the above measurements.
An NQR study confirmed the AF ordering and revealed that essentially all 
the Cu ions are 
involved in the ordering even for the $x = 0.01$ sample \cite{Ohsugi}.
Moreover, suppression of the spin gap 
was suggested from the linear temperature dependence of 
magnetic specific heat above $T_{\rm N}$ for  $x \geq0.02$.
It should be noted here that 
such a linear behavior has been considererd to be characteristic of a gapless 1D AF system 
\cite{Wei}. 
Our data thus implied that there is a finite impurity-induced density of state at $E=0$ 
for $x \geq 0.02$ at least, most probably for $x = 0.01$ also.  
However, it has not been clear how the gap size and the density of states 
change.

In this letter, we report the result of an inelastic neutron 
scattering study on Sr(Cu$_{1-x}$Zn$_{x}$)$_{2}$O$_{3}$ 
($x$ = 0, 0.003, 0.006, 0.01, 0.02 and 0.04).
For the pure ($x$ = 0) sample the opening of a gap of 33 meV (380 K) 
was clearly observed. 
The gap size remained the same
 independent of $x$, whereas the integrated intensity of the magnetic 
scattering, which is proportional to the probability of the 
singlet-triplet excitation over the gap, decreased monotonically until 
it became almost negligible at $x$ = 0.04.

In the case of a polycrystalline sample the scattering law for a 
one-dimensional system is the powder average of the dynamic spin-spin 
correlation function \cite{Mukta}.
\begin{equation}
S({\cal Q}, \omega) = \frac{1}{4\pi {\cal Q}^{2}} T(\omega)|F({\cal Q})|^{2}
\int_{{\bf q}={\bf q}_{\parallel}+{\bf q}_{\perp}, |{\bf q}|={\cal Q}}
S({\bf q}_{\parallel}, {\bf q}_{\perp},  \omega)d {\bf q},
\label{eq:sscor}
\end{equation}
where $F({\cal Q})$ is the ionic form factor, ${\bf q}_{\parallel}$ and ${\bf q}_{\perp}$ 
are the parallel and 
perpendicular projections of the total momentum transfer ${\bf q}$ relative to 
the chain axis, and $T(\omega)$ is the temperature factor 

\begin{equation}
T({\omega}) = 
\left[1 - \exp \left( \frac{-\hbar \omega}{k_{\rm B}T} \right) 
\right]^{-1}.
\label{eq:tempfac}
\end{equation}
Clearly for any given ${\cal Q}$, all values of 4 $\leq {\cal Q}$ will contribute to the 
scattered spectrum.  For dispersive excitations $S({\cal Q}, \omega)$ is 
proportional to the density of states of the dispersion, consequently, 
where there is a singularity in the density of states, such as at a band 
minima, one expects a peak in the scattered intensity. 
For a mode with minima at 
${\bf q}_{\parallel}=q_{1}$ and energy transfer  $\omega=E_{g}$ for example, one 
would expect no contribution to the scattering for  $\omega=E_{g}$, 
${\cal Q} < q_{1}$, 
but a peak in the scattering at $\omega=E_{g}$, ${\cal Q}=q_{1}$ which would persist 
to higher ${\cal Q}$ with the intensity modulated by ${\cal Q}$ and the form factor 
following Eq. (\ref{eq:sscor}).

Powder samples were prepared as described before \cite{Azuma2}.
The neutron-scattering data were collected on the HET 
direct geometry chopper spectrometer at the ISIS pulsed neutron facility 
at the Rutherford Appleton Laboratory \cite{Eccleston}.  
The white pulsed neutron beam was 
monochromated by a Fermi chopper which rotates at frequencies of up to 600 
Hz and was phased to the neutron pulse.  The energy transferred to the 
sample was then calculated from  the time of flight of the scattered 
neutron.  HET is optimized for scattering at low momentum transfers over 
a wide energy range, with banks of detectors at 4 an 2.5m covering the 
angular ranges 2.6$^{{\rm o}}$ to 7$^{{\rm o}}$ and 9$^{{\rm o}}$ to 29$^{{\rm o}}$
 degrees respectively.  
 Two further detector banks at mean scattering angles of 115$^{{\rm o}}$ and
  133$^{{\rm o}}$ are used to 
collect high ${\cal Q}$ data which 
are used for the estimation of the neutron background signal.
About 5 g of a powder sample of each composition was wrapped in a flat Al foil 
sachet and attached 
to the cold finger of a closed cycle refrigerator (CCR).

From the susceptibility data, we anticipated a gap of 36 meV at a momentum 
transfer ${\cal Q}$ = 0.8 \AA$^{-1}$ because the Cu-O-Cu distance along 
the ladder is 3.93 \AA. 
An incident energy of 250 meV was chosen because we were able to reduce 
the ${\cal Q}$ at the energy transfer of 35 meV to 0.93 \AA.
Figure 1(a) shows the data collected on SrCu$_{2}$O$_{3}$ ($x$ = 0.00) at 10 K.
The open circles show the experimental points collected at the mean 
scattering angle of 4.7$^{{\rm o}}$.
The dashed line is an estimation of the scattering from nuclear and the 
single and multiphonon background.  
The latter was estimated from 
data collected in the high-angle detector banks, where the inelastic 
scattering is solely single and multiphonon in origin.  
Two scaling factors for the background were used as free parameters in 
the fit.  
The solid line is the result of fitting to the data after the subtraction 
of this background using the dispersion 
relation for a spin ladder with $J_{\parallel} = J_{\perp}$ predicted by 
Barnes and Riera \cite{Barnes}.  
This fit yields a magnitude of the spin gap of 33 meV in good agreement with 
the value obtained from the susuceptibility measurement.
Figure 2 is the data collected for the same sample at 10, 100 and 200 K.  
Here, the solid lines represent the fit to the data.
The peak was found to broaden with increasing temperature, providing a strong 
evidence that this is magnetic in origin.
Such broadening of the peak had also been reported for a Haldane material 
Y$_{2}$BaNiO$_{5}$ \cite{Darriet}.  
However, such a shift of the peak position as observed for 
that compound was not observed in the present case.

Figure 3 shows the data collected for Sr(Cu$_{1-x}$Zn$_{x}$)$_{2}$O$_{3}$ 
($x = 0, 0.01, 0.04$).  
Surprisingly,  the position of the peak does not shift even at $x = 0.04$.  
Instead, the broadening took place in the data for the doped samples.
The data was analyzed in the same way as described above and the results 
are summarized in Fig. 4.
Figure 4 (a) shows the magnitude of the spin gap  thus estimated, 
which is almost independent of Zn concentration. 
This might seem to be against 
 the assertion that the Zn-substitution makes the ladder gapless.  
On the other 
hand, the integrated intensity of the magnetic scattering between 25 and 
50 meV which corresponds to the possibility of the
singlet-triplet excitation over the spin gap exhibits a different 
behavior.  
As shown in Fig. 4 (b), it decreases monotonically
and the peak is not statistically significant any more at $x = 0.04$.
The change in the low energy excitation could not be observed in this study
because of the strong quasielastic scattering near $E = 0$.
However, on account of the total-sum rule, our result implies that the in-gap density of 
state at $E$ = 0 is finite and grows with increasing $x$. 
The survival of weight at energies of the order of the original spin gap 
had been predicted theoretically \cite{Martins} at a small Zn concentration, however, 
our data revealed that it survives up to a Zn concentration where the $T_{\rm N}$
reaches the maximum .

Present data seem to indicate that impurities introduced in the the singlet 
matrix makes the spins around it alive to a limited extent because of the 
short spin correlation length.  
Remaining singlet pairs are not affected, so the magnitude of the spin gap does not 
change. 
At $x = 0.04$, almost all singlet pairs are destroyed, therefore the spin gap closes.
However, this interpretation cannot explain the $T$-linear magnetic 
specific heat and the fact that all the Cu ions are involved in the AF 
ordering.
Instead, we propose the following picture.
An impurity does not induce a localized magnetic moment
only at its neighboring site, but small staggered spin 
moments appear in the whole region.  
This is the origin of the finite in-gap density of state and the $T$-linear 
magnetic specific heat as well.  
The enhancement of the correlation length 
and the possession of local staggered moments for the sites far from the 
impurities are suggested theoretically \cite{Fukuyama2,Nagaosa,Martins2} 
and experimentally also by an NMR study \cite{Fujiwara}.
Existence of a finite state only at $E$ = 0 
and ${\bf q} = (\pi, \pi)$ seems to be enough for the possession of the staggered 
moments.  
An inelastic neutron scattering study on a single crystal sample is 
required to investigate the ${\bf q}$ dependence of the excitation.  
It is the future work. 
Above the ground state, there still exist a pseudo gap and a triplet excitation 
band.  
Namely, every site has the two exclusive aspects, singlet ground state and 
staggered local moment.
 The in-gap state grows with increasing $x$ until the gap closes 
finally around $x$=0.04.  

The induced moments get 
magnetically ordered at low temperatures because of the small three
dimensionality. 
However, the ordered moment is quite small because of the
spin frustration at the interface between the ladders.  
This is 
consistent with the result of the NQR study which revealed that the ordered 
moment is as small as 0.01 $\mu_{\rm B}$ \cite{Ohsugi}. 
The $T_{\rm N}$ first increases with increasing $x$, 
while it next descends because the impurity 
ions work to cut the correlation.

In summary, we have performed an inelastic neutron scattering study on 
Sr(Cu$_{1-x}$Zn$_{x}$)$_{2}$O$_{3}$ ($x \leq 0.04$) to investigate 
nonmagnetic impurity effects on the 2-leg spin ladder system.
The existence of the spin gap of 33 meV (380 K) was confirmed in the pure 
sample.  Its magnitude was found to be independent of Zn concentration, 
while the integrated intensity of the magnetic scattering which 
corresponds to the singlet-triplet excitation decreases monotonically.
These data suggest the finite in-gap density of state at $E$=0
 which leads to the $T$-linear term of the specific heat.

This work was partly supported by a Grant-in Aid for 
Scientific Research on Priority Areas, ``Anomalous metallic state near the 
Mott transition'', of Ministry of Education, 
Science and Culture, Japan and CREST (Core Research for Evolutional 
Science and Technology) of Japan Science and Technology Corporation (JST).

\begin{figure}
\caption {Scattering from SrCu$_{2}$O$_{3}$ with an incident energy of 250 
meV at a mean scattering angle of 4.7 $^{{\rm o}}$
collected at 10 K. The solid line is the magnetic component and the 
dashed line is the nuclear and single and multiphonon scattering.}
\label { fig. 1}
\end{figure}
\begin{figure}
\caption {Scattering from SrCu$_{2}$O$_{3}$ at 10, 100 and 200 K.  The 
solid lines represent the fit to the data.}
\label { fig. 2}
\end{figure}
\begin{figure}
\caption {Scattering from Sr(Cu$_{1-x}$Zn$_{x}$)$_{2}$O$_{3}$ 
($x$ = 0, 0.01 and 0.04) collected at 10K}
\label { fig. 3}
\end{figure}
\begin{figure}
\caption {(a) Magnitude of the spin gap of 
Sr(Cu$_{1-x}$Zn$_{x}$)$_{2}$O$_{3}$ ($x$ = 0, 0.003, 0.006,0.01, 0.02 and 0.04) estimated
 from the scattering data. 
 (b) Integrated intensity of magnetic scattering observed between 25 and 50 meV.}
\label{fig. 4}
\end{figure}
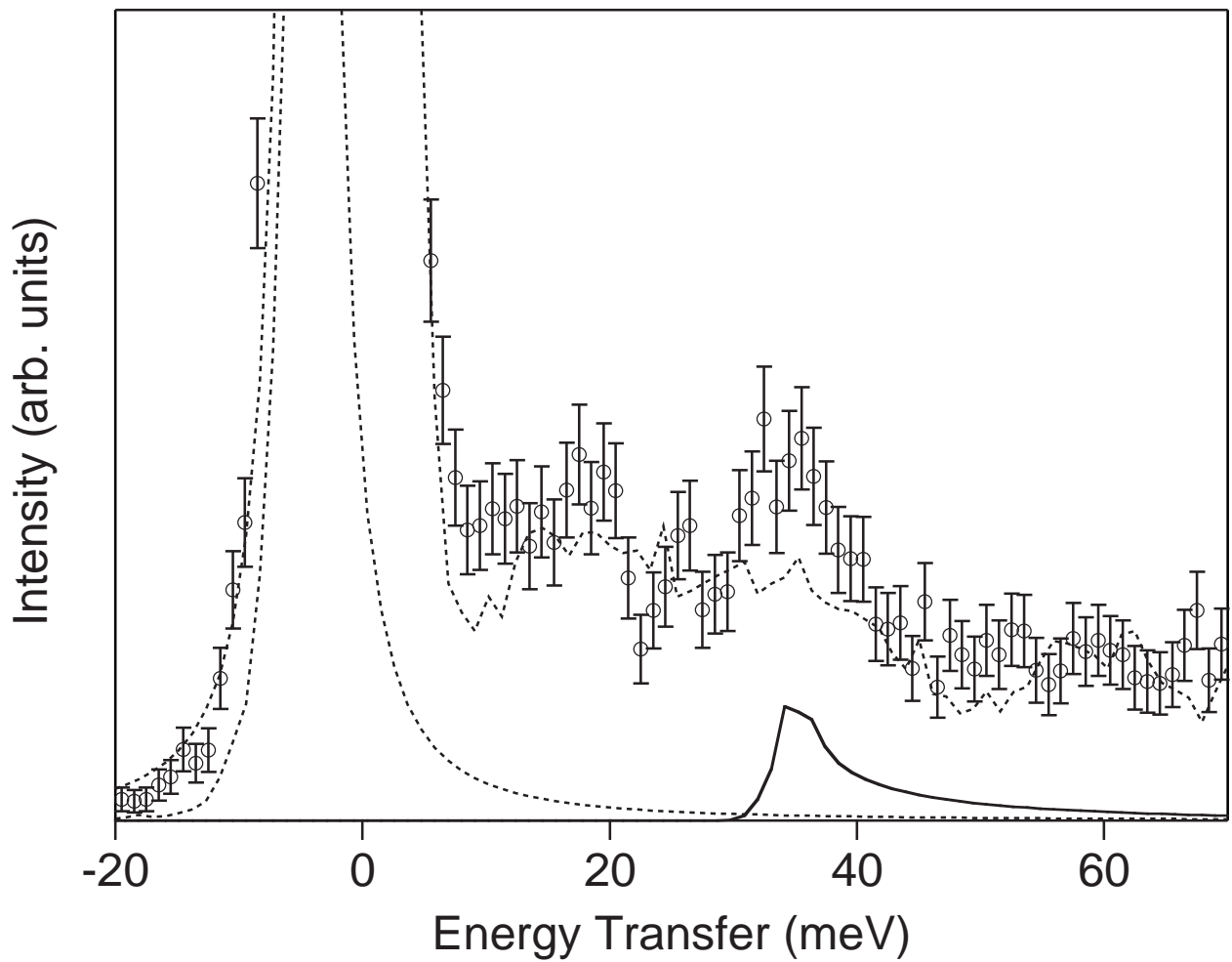
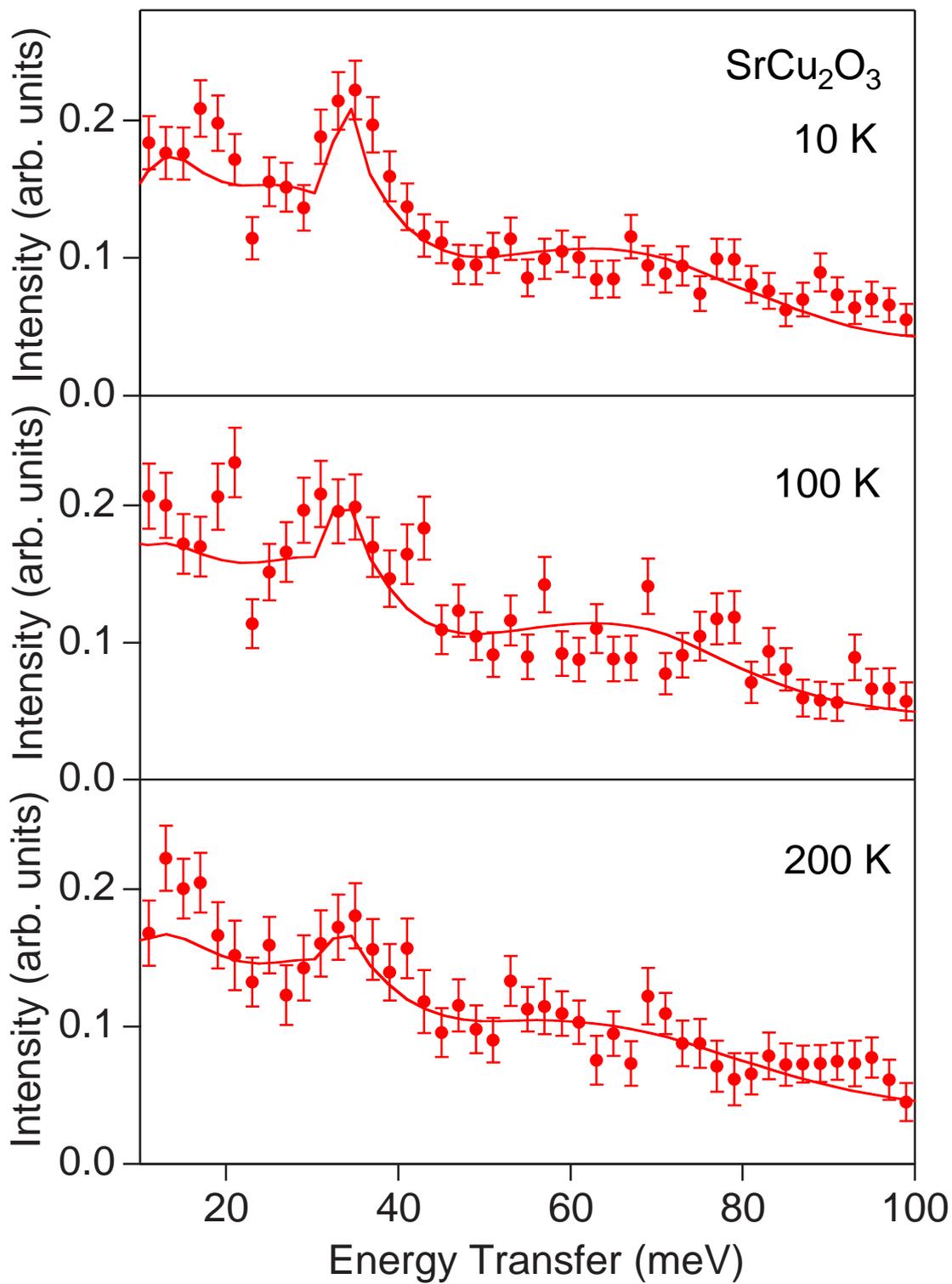
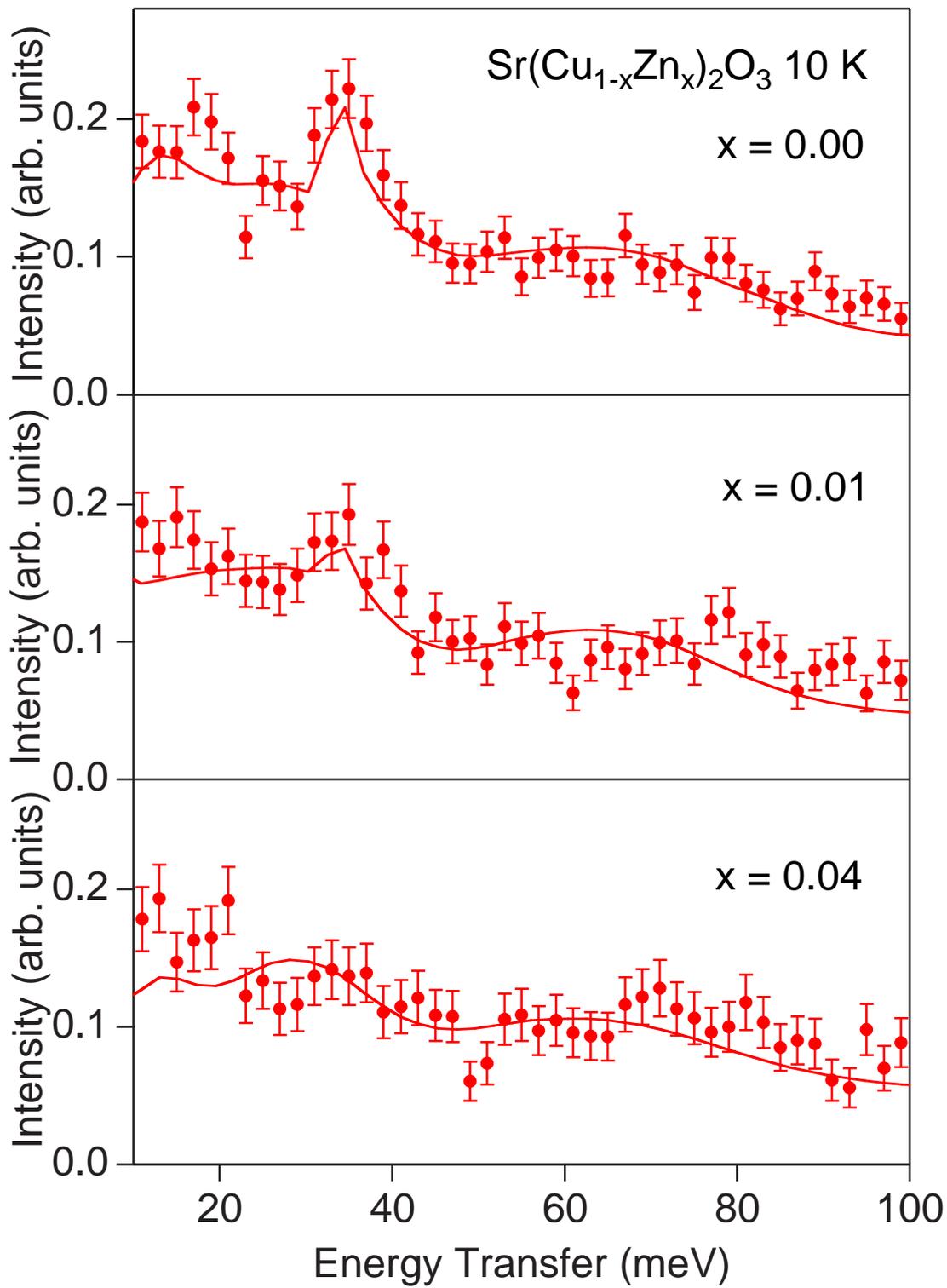
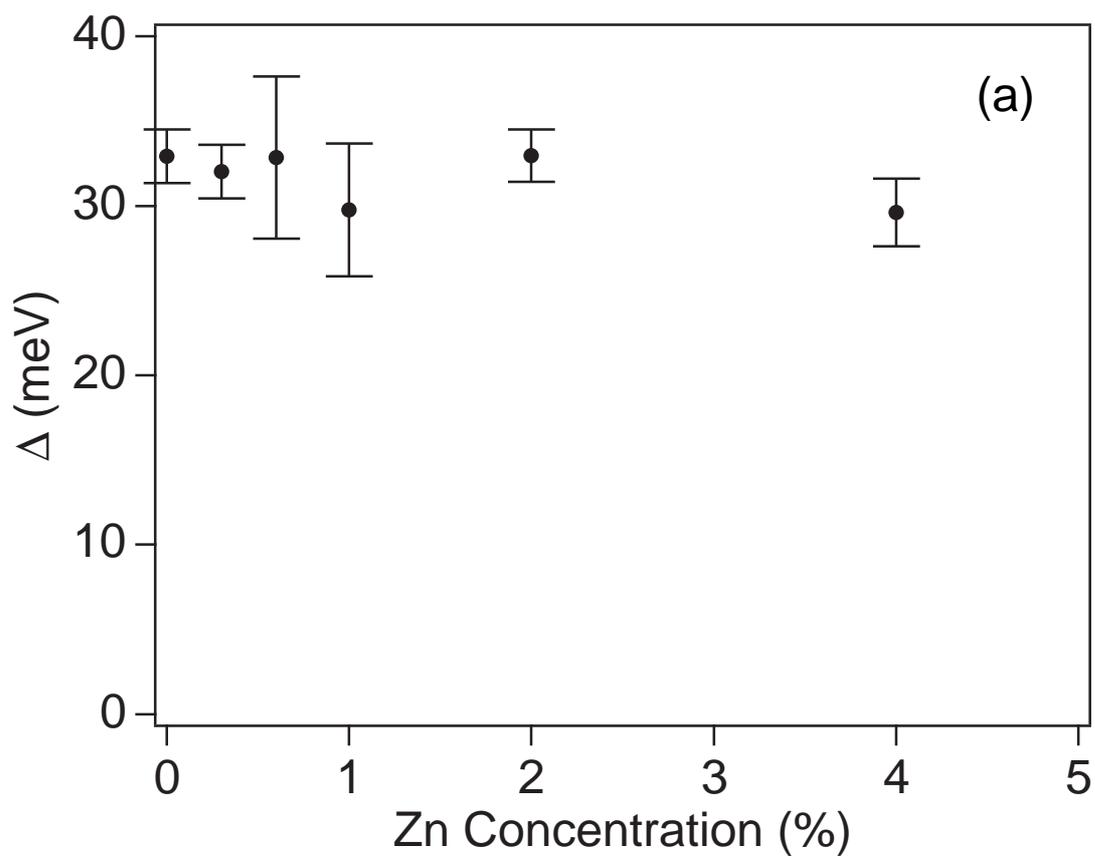
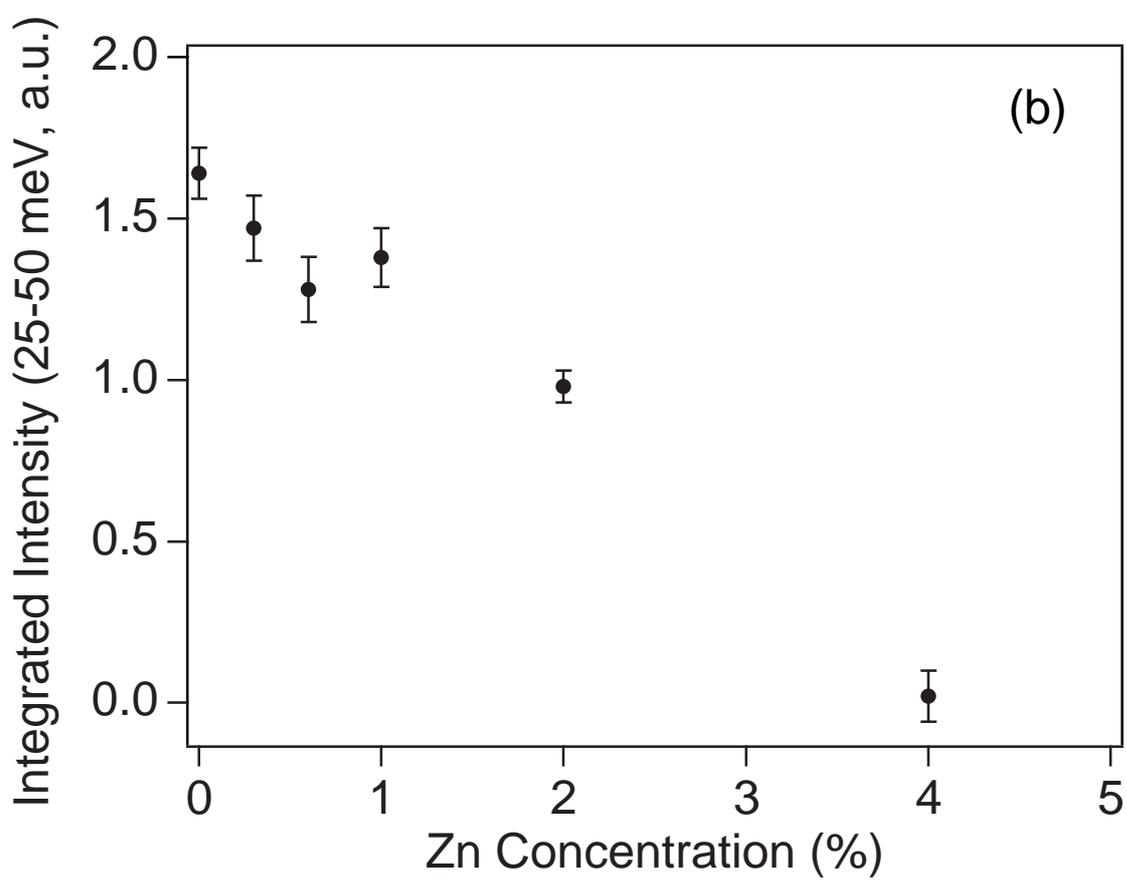
\end{document}